\documentclass[%
 reprint,
superscriptaddress,
 amsmath,amssymb,
 aps,
prl,
floatfix,
]{revtex4-2}

\usepackage{graphicx}
\usepackage{dcolumn}
\usepackage{bm}
\usepackage{siunitx}
\usepackage{xcolor}
\usepackage{lmodern}
\usepackage[hidelinks]{hyperref}
\usepackage{orcidlink}
\usepackage{calc}

\usepackage{changes}
\definechangesauthor[name={Martijn}, color=magenta]{mlr}
\definechangesauthor[name={Julian}, color=orange]{jcb}
\definechangesauthor[name={Emily}, color=blue]{evk}

\begin{document}


\newcommand\ARCNL{Advanced Research Center for Nanolithography (ARCNL), Science Park 106, 1098 XG Amsterdam, The Netherlands.}

\newcommand\VU{Department of Physics and Astronomy, and LaserLaB, Vrije Universiteit Amsterdam, De Boelelaan 1100, 1081 HV Amsterdam, The Netherlands.}

\title{Resolving the Blueshift in Calculations of \\ the EUV Spectrum of Multiply Charged Tin Ions}

\begin{abstract}
We report \emph{ab initio} relativistic calculations on the complex open shell Sn$^{12+}$ highly charged ion, a prototypical plasma ion relevant for extreme ultraviolet (EUV) nanolithography. Previous calculations of EUV emissivity in tin plasmas consistently generate a spectrum in which the region of peak emissivity is blueshifted relative to experiment. By optimising our numerical methods to take full advantage of modern, high-performance CPU architectures, we are able to completely saturate the configuration interaction within the $n=4$ shell. Coupled with a thorough treatment of core-valence correlation we resolve the blueshift, finding a surprisingly large influence of highly excited states on the spectrum that is dominated by multiply excited states.
\end{abstract}

\author{M.L. Reitsma\orcidlink{0000-0002-8255-7480}}
\affiliation{\ARCNL}
\author{J. Sheil\orcidlink{0000-0003-3393-9658}}
\affiliation{\ARCNL}
\affiliation{\VU}
\author{O.O. Versolato\orcidlink{0000-0003-3852-5227}}
\affiliation{\ARCNL}
\affiliation{\VU}

    \makeatletter
    \let\tmpaffiliation\affiliation
    \let\tmpauthor\author
    \let\tmpabstract\frontmatter@abstract@produce
    \let\frontmatter@abstract@produce\relax
    \let\frontmatter@finalspace\relax
    \maketitle
    
    \def\frontmatter@finalspace{\addvspace{18\p@}}
    \let\maketitle\frontmatter@maketitle
    \let\affiliation\tmpaffiliation
    \let\author\tmpauthor
    \let\frontmatter@abstract@produce\tmpabstract
    \let\frontmatter@title@produce\relax
    \makeatother

\author{E. V. Kahl\orcidlink{0000-0003-3923-7120}}
\affiliation{Pawsey Supercomputing Research Centre, 1 Bryce Avenue, Kensington, 6151, WA, Australia}
\author{J. C. Berengut\orcidlink{0000-0002-7366-1091}}
\affiliation{School of Physics, University of New South Wales, Sydney NSW 2052, Australia}

\date{\today}

\maketitle


\textit{Introduction}.--- Laser-produced tin plasmas are the source of 13.5\,nm-wavelength radiation used in state-of-the-art nanolithography for computer chip production~\cite{Kazazis2024,Wagner2010,Waldrop2016,Banine2011,Fomenkov2017,Kawasuji2017,Versolato2019,Zhu2025}. This radiation has its origin in millions of atomic transitions between multiply excited states of Sn$^{11+}$ $ - $ Sn$^{14+}$ ions~\cite{torretti2020prominent,Sheil2021,sasaki2024effect,yan2024theoretical}. The resulting extreme ultraviolet (EUV) spectrum exhibits an intense, $\sim$0.6\,nm-wide emission feature peaked near a wavelength of 13.5\,nm~\cite{Fujioka2005opacity,torretti2020prominent}. Predicting the amount of EUV radiation produced by the plasma (as well as energy that is funneled into other channels) remains a very challenging task for academics and industrialists alike~\cite{Sheil2023}. Accuracy in the line positions is essential; only radiation in a narrow 2\%, or 0.27\,nm, band is reflected by Mo/Si multilayer mirrors in industrial machines~\cite{Bajt2002} and utilized in EUV lithography. This fact, coupled with the need for extensive coverage of multiply excited states~\cite{torretti2020prominent,Sheil2021,sasaki2024effect,yan2024theoretical}, significantly complicates the generation of opacities and emissivities for interpreting experimental spectra~\cite{Sheil2023} and for use in radiation-hydrodynamic simulations, the toolkit-of-choice for EUV source design and optimization~\cite{Purvis2016,Scott_laser_plasma,Su2017,Basko2016}. 

Despite years of intense research on the atomic physics of multiply charged tin ions, accurate predictions of energy levels and atomic spectra using \emph{ab initio} atomic structure methods have remained out of reach. This is especially true for Sn ions with multiple electrons in the $ 4d $ subshell, excitations from which produce a dense structure of strongly interacting configurations.  To date, semi-empirical corrections have proven essential for bringing theory into close agreement with experiment. For calculations performed with the semi-relativistic Cowan code~\cite{Cowan1981}, these corrections take the form of ``scale factors'' that pre-multiply Slater integrals that appear in the Hamiltonian matrix to take into account the ``infinity of small perturbations''~\cite{Cowan1981} that arise from configurations not included in the CI calculation. One typically adjusts these scale factors to values $ < 1 $ to bring theoretical and experimentally derived energy levels into agreement; scaling factors appropriate for Sn ions are $\sim$0.85 for singly excited states~\cite{Churilov2006SnIX--SnXII,Churilov2006SnVIII,Churilov2006SnXIII--XV,Tolstikhina2006ATOMICDATA, Ryabtsev2008SnXIV,darcy2009transitions,darcy2009identification,kieft2005comparison,Windberger2016,Torretti2017,Ryabtsev2017,Ryabtsev2019, Su2017,Colgan2017,scheers2020euv} and $\sim$0.75 for multiply excited configurations. This procedure of reducing the strength of valence-valence interactions essentially mimics the effect of the screening of such interactions by core electrons~\cite{Dzuba1996,Dzuba2008}. 

Relativistic codes such as HULLAC~\cite{HULLAC_code}, FAC~\cite{FAC_code} or the GRASP family of codes~\cite{GRASP92_code,GRASP2K_code,GRASP2018} do not implement such semi-empirical capabilities. \emph{Ab initio} calculations performed with these codes generally overestimate excited level energies in Sn ions, resulting in calculated spectra that peak near a wavelength of $\sim$13\,nm~\cite{Koike2007,Sasaki2010,Zeng2010,scheers2020euv}. In the present work, we refer to this $\sim$0.5\,nm offset between experiment and \emph{ab initio} theory as a ``blueshift''. Importantly, this blueshift also pertains to transitions between multiply excited states~\cite{torretti2020prominent,Sheil2021,sasaki2024effect,yan2024theoretical}. It is significantly larger than the 0.27\,nm acceptance bandwidth of the industry-standard mirrors, thus precluding predictive modeling.

The inability of \emph{ab initio} atomic structure calculations to reproduce experimental spectra lies in the presence of strong configuration interaction (CI) in this system. For one, the radial wavefunctions of the $ 4p, 4d $ and $ 4f $ electrons strongly overlap~\cite{Kucas2009,OSullivan2015}. Second, because the binding energies are similar for these electrons~\cite{Koike2007}, configurations formed by exciting electrons with different angular momenta (keeping $ n = 4 $ fixed) overlap in energy space. This leads to strong mixing between levels of the singly excited $ 4p^{5}4d^{n+1} $ and $ 4p^{6}4d^{n-1}4f $ configurations~\cite{Karazija2006,Dortan2007,Karazija2013} as well as multiply excited configurations~\cite{torretti2020prominent}. Transitions from these strongly mixed configurations give rise to very complicated spectra, where CI results in a narrowing and redistribution of oscillator strength towards the higher-energy side of the transition array~\cite{Bauche1987,Mandelbaum1987,Sullivan1999,Karazija2006,Dortan2007,Kucas2009, OSullivan2015}.

In this work, we present high-accuracy, large calculations of the structures and spectra of Sn$^{12+}$, a prototypical plasma ion that contributes strongly to EUV emission~\cite{torretti2020prominent},  using the CI + many-body perturbation theory (MBPT) approach. With this method, we have been able to include all possible interactions between configurations arising from permutations among electrons in the $ n = 4 $ shell: we \textit{fully saturate} the CI space. This is made possible by the CI+MBPT approach, where interactions between valence electrons (i.e. valence-valence correlations) are computed using CI calculations, and core-valence correlations are computed via MBPT methods. This partitioning is key to enabling a complete description of the interacting space. We show that transition energies between multiply excited states can only be described accurately by including a high number of excitations within the $n=4$ shell, more specifically, the configuration interaction between states that have up to six electrons removed from the $4p$ shell has to be included. In addition, we show that core-valence correlation is crucial for accurately calculating both singly- and multiply-excited states, while higher-order relativistic contributions are smaller. Crucially, our \emph{ab initio} calculations align exactly at 13.5\,nm, resolving the longstanding EUV blueshift anomaly.

\textit{Method}.---
We use the particle-hole CI+MBPT (configuration interaction combined with many body perturbation theory) method~\cite{berengut16pra} implemented in the AMBiT program~\cite{Kahl2019}, which extends the CI+MBPT method~\cite{Dzuba1996} to allow for valence holes to be treated in CI. 
The Dirac-Coulomb Hamiltonian for our calculations is (in atomic units $\hbar = m_e = e = 1$)
\begin{equation}
    H_{DC} = \sum_i h_D(i) + \sum_{i<j} 1/r_{ij},
\end{equation}
where $h_D$ is the one-electron Dirac Hamiltonian,
\begin{equation}
    h_{D}(i)=c\, \boldsymbol \alpha_{i}\cdot \mathbf{p}_{i}+c^{2}(\beta _{i}-1)+V_\text{nuc}(i).
\end{equation}
Here, $\boldsymbol \alpha$ and $\beta$ are the four-dimensional Dirac matrices, and a Fermi charge distribution~\cite{Visscher97} is used for the nuclear potential $V_\text{nuc}$. Additionally, we calculate the magnitude of the frequency-independent Breit interaction, self-energy, and (Uehling) vacuum polarization effects using an effective potential~\cite{flambaum2005radiative,ginges2016atomic,ginges2016qed} in separate calculations.

We begin with a self-consistent Dirac-Hartree-Fock (DHF) calculation for $N=38$ electrons in the V$^{N}$ approximation corresponding to the configuration-averaged ground state [Kr]~$4d^2$. We then generate our orbital basis by diagonalizing a complete set of B-splines over the DHF potential. The $4p$, $4d$, and $4f$ shells are treated as valence and correlated in our CI expansion as explained below.

In the CI+MBPT method, core-valence excitations are treated at second order in the residual Coulomb interaction by modifying the one and two-particle integrals in the CI procedure~\cite{Dzuba1996}. Effective three-body operators are separately added to the Hamiltonian matrix elements. The core-valence diagrams include virtual orbitals up to $30spdfghi$ (i.e. $n\leq30$ and $l\leq6$). We also performed small-scale tests up to $n=35$ and $l=7$ to confirm that the MBPT contributions are well-converged at this level. Using AMBiT's implementation of the particle-hole CI+MBPT theory, we place the Fermi level above the $4p$ shell and treat $4p$ vacancies as holes, which reduces the size of subtraction diagrams compared to treating $4p$ electrons as valence particles. We have also performed calculations starting from the alternative closed-shell $V^{N-2}$ potential, for which there are no subtraction diagrams in the particle-hole CI+MBPT method, and found that our results are stable with respect to the initial DHF potential.

\textit{CI expansion strategy}.---
\begin{figure*}[htbp]
    \centering
    \includegraphics[width=1.0\linewidth]{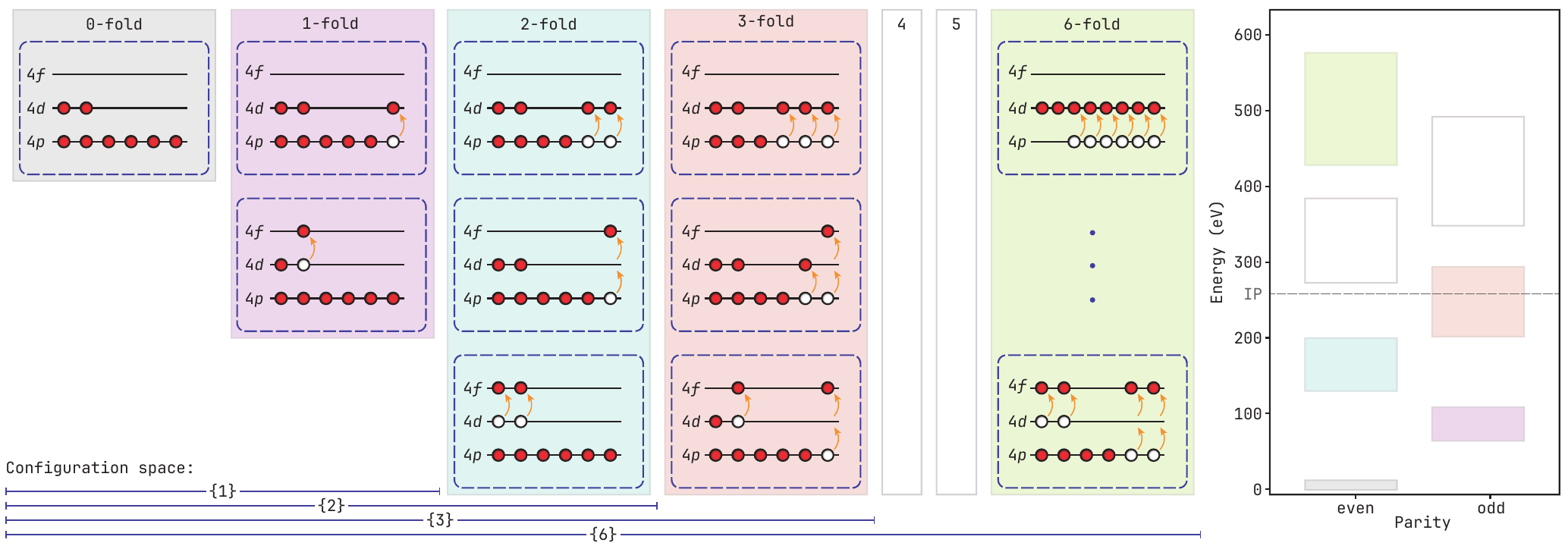}
    \caption{Electronic configurations of Sn$^{12+}$ for 1- through 6-fold excitations from $4p^6$ to the $4d$ and $4f$ orbitals (left), and the level structure arising from such configurations (right), where each shaded region encloses the energy range of a corresponding $m$-fold.}
    \label{fig:folds_blobs}
\end{figure*}
We expand the CI configuration space using a systematic procedure within the $n=4$ shell, akin to a superconfiguration description of atomic structures, which is illustrated in Fig.~\ref{fig:folds_blobs}. The ground state manifold of Sn$^{12+}$ is [Zn]~$4p^6 4d^2$, which we call the 0-fold configuration. Taking single-electron excitations from the $4p$ to the $4d$ subshells, or from the $4d$ into the $4f$, produces the first band of excited state energy levels: the 1-fold configuration set. Next, the 2-fold set contains configurations formed from the ground state manifold by the excitation of two electrons from the $4p$ into the $4d$ orbitals, or from the $4d$ into the $4f$ orbitals, or a single electron excited from the $4p$ into the $4f$ orbital. This procedure continues up to the 6-fold configuration set where six electrons are removed from the $4p^6 4d^2$ configuration and distributed over the $4d$ and $4f$ orbitals. As such, each $m$-fold corresponds to all ways of distributing electrons using n excitations.

CI calculations performed on a configuration space consisting of one or more of these configuration sets are designated by \{n\} and include all configuration sets up to the $m$-fold set. For example, the \{3\} CI configuration space includes all configurations belonging to the 0-fold, 1-fold, 2-fold and 3-fold sets. This naming scheme should not be confused with the names of different transition sets; The transition set between the 1-fold and 0-fold states will be designated 1$\rightarrow$0, while 2$\rightarrow$1 and 3$\rightarrow$2 correspond to the transition sets between the 2-fold and 1-fold, and 3-fold and 2-fold states, respectively.

In total, the full \{6\} CI space includes 22 configurations in the $n=4$ shell, corresponding to 115\,715 levels. We analyze the electric dipole ($E1$) transitions between states up to the 3-fold set, which adds up to 1\,613\,349 $E1$ transitions between the 5\,294 levels.

The efficient calculation of these levels and transition strengths has been enabled by two main improvements to the AMBiT code. Firstly, one of the slowest parts of the CI+MBPT method is the generation of the CI matrix. Previously in AMBiT, all configuration state functions (CSFs) corresponding to a relativistic configuration were grouped together in the matrix, and the time taken to generate matrix elements between different groups have an extremely skewed distribution. Uneven load-balancing between threads results in wasted compute-time, which severely limits parallel scalability on systems with large numbers of CPU cores such as the Dutch national supercomputer Snellius, which we have used for all calculations presented in this work. We have optimized the workload balance over all threads using a mix of static, physics-informed load balancing with MPI, and dynamic load balancing with OpenMP tasks to minimize wasted threads, giving a speedup of ${\sim} 3\times$ compared to the naive load-balancing scheme.

We have also improved the performance of calculating matrix elements between the CSF groups by re-expressing the algorithm in terms of blocked, dense matrix operations. This change is responsible for a speedup of ${\sim} 5\times$ due to improved data localization and CPU cache utilization - both are critical to achieving peak performance on high-performance CPU architectures.

Also important to these performance gains is the use of modern, cache-aware C++ data-structures in the \textit{abseil} library~\cite{Benzaquen2017}. AMBiT makes heavy use of hash tables and search trees, but the C++ standard template library (STL) implementation of these data structures (\texttt{std::map} and \texttt{std::unordered\_map}, respectively) have large amounts of indirection and ``pointer-chasing'' which limits their ability to make use of the large caches in modern CPU architectures. The flat hash table (referred to as a \textit{Swiss Table}) and B-Tree data structures in \textit{abseil} are designed to optimize both algorithmic efficiency and cache utilization for a variety of access patterns~\cite{Benzaquen2017, Graefe2011}. We find that using cache-aware data structures results in a further speedup of ${\sim} 2\times$ compared to the baseline prior to this work. In total, both the CI matrix generation and the calculation of transition strengths have been sped up by more than an order of magnitude. The latter would have been computationally intractable without these software engineering advances.

\textit{Results and discussion}.---
In order to quantify the magnitude of the different transition bands' contribution to the EUV emission, we quantify the weighted mean $\mu$ of a calculated transition set by weighting each transition energy $\Delta E_i = E_i^\text{upper} - E_i^\text{lower}$ by its degeneracy-weighted transition probability $g_iA_i$ as 
\begin{equation}
    \mu_E = \frac{\sum_i g_iA_i \Delta E_i}{\sum_i g_iA_i}\ \text{or}\ \mu_\lambda = \frac{hc}{\mu_E},
\end{equation}
in terms of energy and wavelength, respectively.

\begin{table}[tb]
    \caption{Calculated degeneracy-weighted mean energies of $n=4$ transitions in Sn$^{12+}$.}
    \label{tab:ci_and_corrections}
    \centering
    \begin{tabular*}{\columnwidth}{l  @{\extracolsep{\stretch{1}}}*{3}{c}@{}}
        \hline
        \hline
         Level of theory & \multicolumn{3}{c}{Transition energy (eV)} \\
         \cline{2-4}
         & $1\rightarrow0$ & $2\rightarrow1$ & $3\rightarrow2$  \\
        \hline
                  CI \{1\} & 96.39  & --    & -- \\
        \phantom{CI }\{2\} & 100.80 & 94.80 & -- \\
        \phantom{CI }\{3\} & 94.68  & 98.87 & 93.14 \\
        \phantom{CI }\{4\} & 95.16  & 93.52 & 96.89 \\
        \phantom{CI }\{5\} & 94.72  & 93.87 & 92.20 \\
        \phantom{CI }\{6\} & 94.73  & 93.55 & 92.46 \\
        \hline
        $\Delta$MBPT     & -2.04  & -2.70 & -3.16 \\
        $\Delta$Breit    & -0.17  & -0.15 & -0.14 \\
        $\Delta$QED      & -0.003 & -0.001& -0.001  \\
        \hline
        Total (eV)       & 92.52  & 90.69 & 89.16 \\
        Total (nm)       & 13.40  & 13.67 & 13.91 \\
         \hline
         \hline
    \end{tabular*}
\end{table}
The results of the calculations of the mean energy for each transition set at different levels of CI saturation \{1\} through \{6\} is presented in Table~\ref{tab:ci_and_corrections} alongside the additional corrections for core-valence correlation ($\Delta$MBPT), Breit interaction ($\Delta$Breit) and self-energy and vacuum polarization effects ($\Delta$QED).

The core-valence correlation at second order in MBPT was extracted from a CI+MBPT calculation using the \{6\} CI space. The QED and Breit corrections were also calculated at the \{6\} CI space. Prominently, core-valence correlations ($\Delta$MBPT) account for the largest corrections. These shift the 1$\rightarrow$0 mean transition energies by -2.04\,eV, while the higher 2$\rightarrow$1 and 3$\rightarrow$2 transitions are even more strongly affected, with corrections of -2.70\,eV and -3.15\,eV, respectively. For the $n=4$ CI saturation, let us take \{3\} as the reference point, as this contains approximately the same $n=4$ configurations that were included in calculations in other literature. The mean energy of the 1$\rightarrow$0 set is relatively well-converged at \{3\}, differing from \{6\} by only 0.05\,eV. For the other two sets the CI space is not well converged at \{3\}, as we note that the 2$\rightarrow$1 mean is lowered by 5.32\,eV and 3$\rightarrow$2 by 0.68\,eV when going from \{3\} to \{6\}. The Breit corrections are small, and the QED corrections are negligible even in this highly charged ion, because the transitions do not involve the $s$-wave orbitals and the atomic number of tin is not particularly high.

The convergence of the transition energies as a function of the CI space requires a more detailed discussion.
\begin{figure}[bt]
    \centering
    \includegraphics[width=\linewidth]{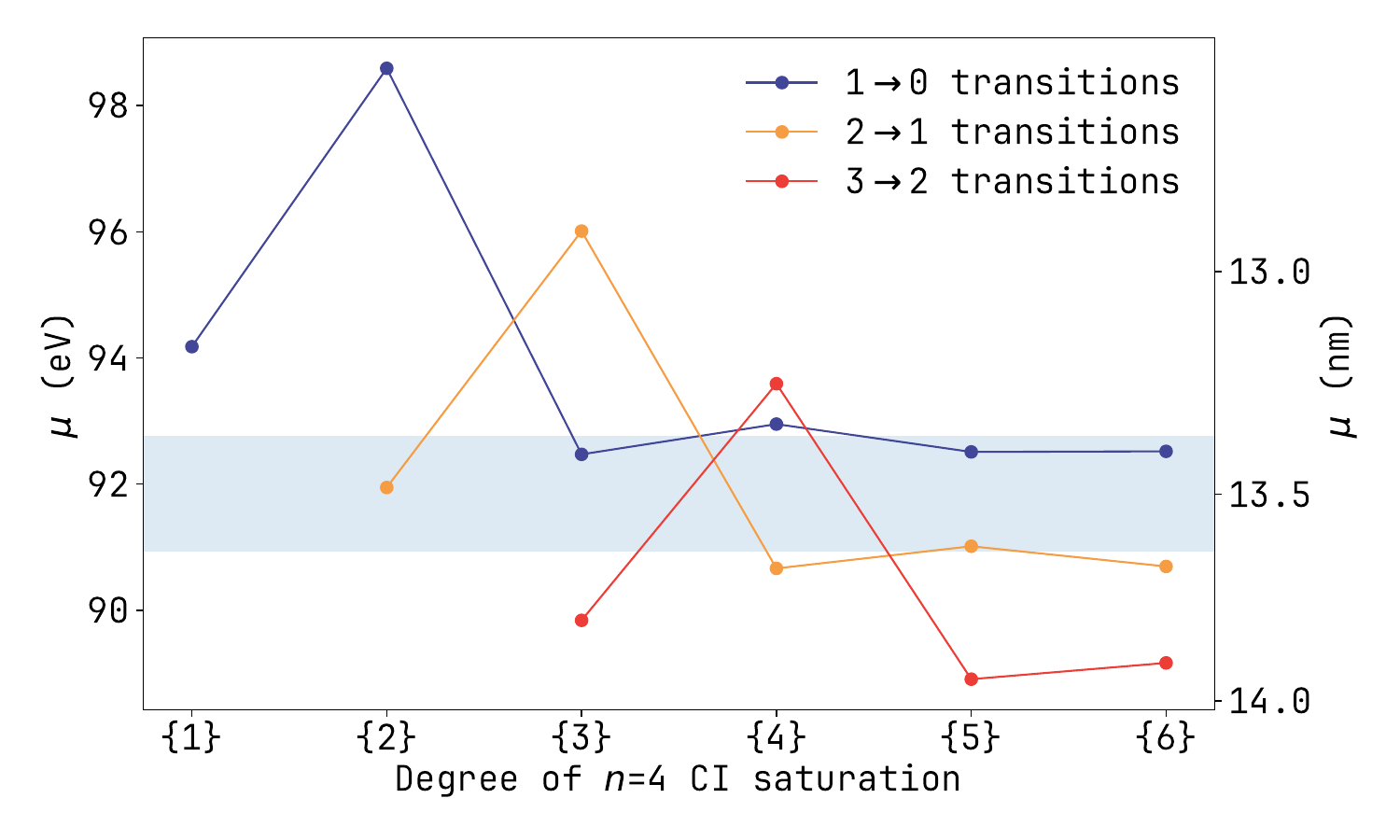}
    \caption{Convergence of the mean degeneracy-weighted transition energy with respect to the CI space saturation for each transition set, including $\Delta$MBPT, $\Delta$Breit and $\Delta$QED. The shaded region indicates a 2\%-bandwidth around 13.5\,nm.}
    \label{fig:alps_abs}
\end{figure}
Fig.~\ref{fig:alps_abs} shows the mean of the transition sets at different levels of CI corrected by $\Delta$MBPT, $\Delta$Breit and $\Delta$QED. A distinct up-down pattern is revealed, which originates from the manner in which the CI space is built up. As shown on the right of Fig.~\ref{fig:folds_blobs}, the energy levels of any two consecutive folds are opposite in parity, and $E1$ transitions are only allowed between those opposite-parity states. In contrast, configuration mixing is only possible between same-parity states. Therefore, when increasing the CI space by one step, the newly added states mix only with the lower or with the upper states of the transition, lowering their energy relative to the other, strongly changing the transition energy. In the next step, new opposite parity states are added and the situation is reversed, largely reversing the change in transition energy.

We find that the mean energy of the $1\rightarrow0$ transitions are shifted up by over 4~eV when going from CI space \{1\} to \{2\}, and then down again by about 6~eV when going to the \{3\} space. Interestingly, the amplitudes of these shifts are almost identical for the $2\rightarrow1$ and $3\rightarrow2$ transitions, where these steps occur in the \{2\}--\{3\}--\{4\} and \{3\}--\{4\}--\{5\} range respectively. Convergence at $\sim$0.3~eV, is finally reached between space \{5\} and \{6\}. This means that the convergence of the CI space for the lower excited states (i.e. corresponding to $1\rightarrow0$) does not imply convergence for the higher excited states. Moreover, to accurately predict transition energies between multiply-excited states, it is necessary to include the full \{6\} space that includes all configurations from the 0-fold up to the 6-fold excitations from the $n=4$ ground state.

The calculation of the 1$\rightarrow$0 mean transition energy can be directly compared to the mean transition energy of known experimental lines for Sn$^{12+}$ as published in Ref.~\cite{Churilov2006SnXIII--XV}. Using their reported transition energies and rates for $n=4$ lines, we determine the corresponding $\mu_{1\rightarrow0}^\text{expt}=13.37$\,nm, in close agreement with our result of $\mu_{1\rightarrow0}^\text{theory}=13.40$\,nm. It should be noted that we do not expect to fully predict the experimental spectra in this work, but rather to demonstrate the crucial core-valence and valence-valence correlation contributions that explain the large discrepancies in previously calculated spectra. Thus, the contributions that were described in detail in previous literature (e.g., Ref.~\cite{torretti2020prominent}) were not considered for our main result. Nevertheless, the close agreement of our prediction with experiment demonstrates that the CI saturation of the $n=4$ shell and core correlation are dominant and cannot be ignored for an accurate description of the electronic structure around the 13.5\,nm wavelength. 

A complete description of the Sn$^{12+}$ spectrum should include configuration mixing from the $4s$ shell and lines from the $n=5$ shells. We have performed additional calculations to determine the effect of these configurations, and found that $\mu$ changes by 0.03\,nm for 1$\rightarrow$0 (-0.03\,nm for 2$\rightarrow$1 and 3$\rightarrow$2). This is insignificant compared to the contributions of core-correlation and CI saturation within the $n=4$ shell.

\begin{figure*}[htbp]
    \centering
    \includegraphics[width=\linewidth]{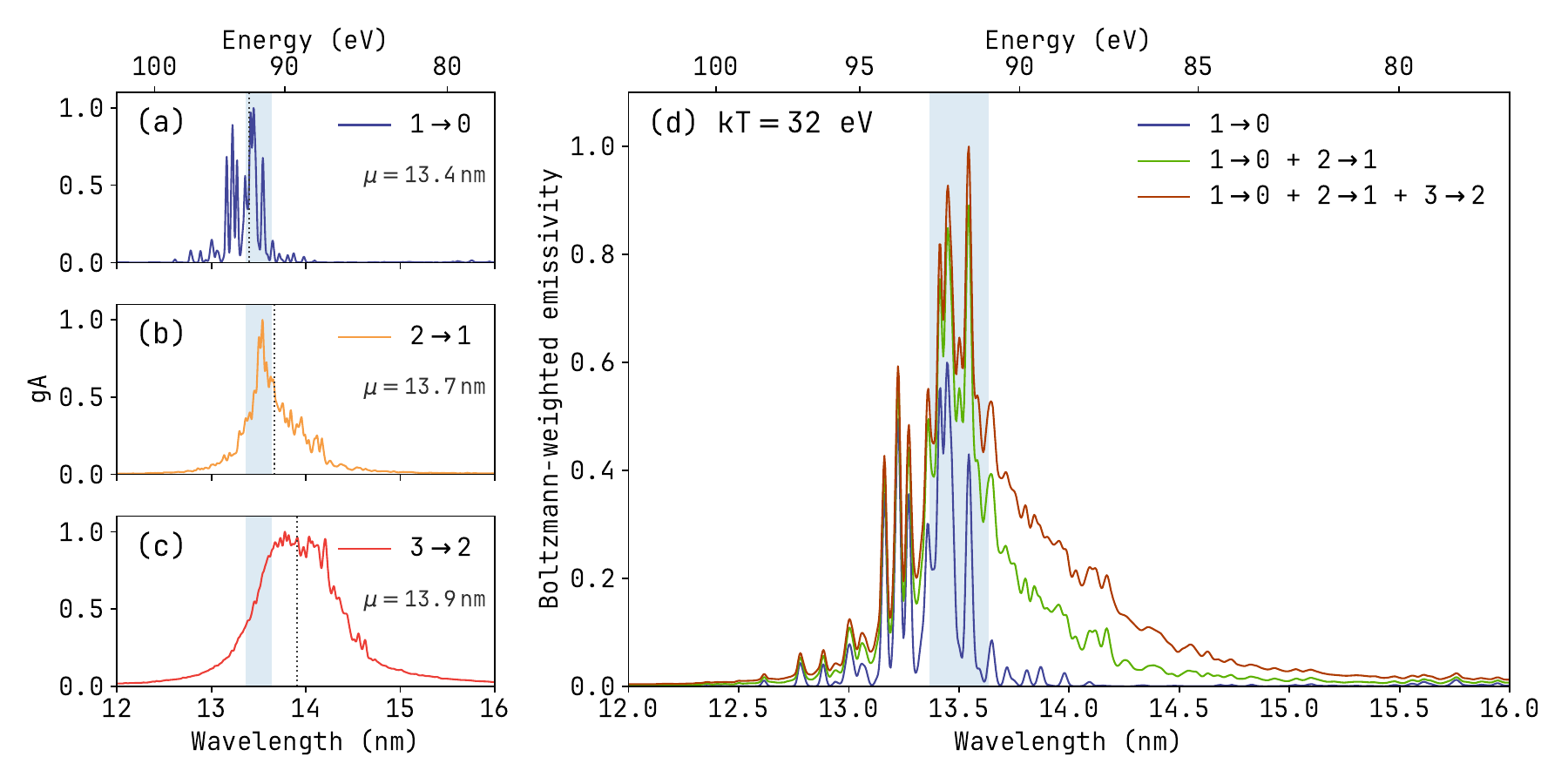}
    \caption{Calculated transition distributions of Sn$^{12+}$, normalized to the maximum of each distribution. (a--c) Degeneracy-weighted transition probabilities $gA$ for each transition set. A dotted vertical line indicates the mean of each distribution. (d) Boltzmann-weighted emissivity $gA\exp(-E_i^\text{upper}/kT)$ for cumulative combinations of transition sets, at a temperature of $kT=32$\,eV. Each spectrum is convoluted by a Gaussian profile to improve visibility. The shaded regions indicate a 2\%-bandwidth around 13.5\,nm.}
    \label{fig:strength}
\end{figure*}
Fig.~\ref{fig:strength}(a-c) shows the distributions of the degeneracy-weighted transition probabilities $gA$ for each transition set, which include $\Delta$MBPT, $\Delta$Breit and $\Delta$QED corrections. From these distributions we obtain a Boltzmann-weighted emissivity by multiplying each $gA$ by the Boltzmann population factor $\exp(-E_i^\text{upper}/kT)$, and the resulting emissivity spectra are shown in Fig.~\ref{fig:strength}(d). Here, we show consecutive sums of emission spectra of each transition set, such that the overarching spectrum consists of the sum of the emissivities of all three transition sets. We set the temperature at $kT=32$\,eV in accordance with the tin-EUV plasma temperature of Ref.~\cite{torretti2020prominent}. The shaded region represents the industry-relevant 2\% (or 0.27\,nm) reflectivity bandwidth around 13.5\,nm. 

Looking only at the degeneracy-weighted transition probabilities, we see that especially for the 2$\rightarrow$1 and 3$\rightarrow$2 sets, a large number of lines are located outside the 13.5\,nm region. The Boltzmann-weighted emissivity shows, however, that at typical plasma conditions the electronic states responsible for the 13.5\,nm emission are predominantly populated, leading to a narrow peak that is centered very precisely at the important 13.5\,nm region. This is the key outcome of our work: Without the necessity for any empirical adjustments, our thorough \emph{ab initio} calculations can accurately position the peak of the tin-EUV spectrum at the wavelength that coincides with experimental observations. Thus, the long standing issue of the EUV blueshift is resolved through complete CI saturation and the thorough inclusion of core correlation.

\textit{Conclusion}.---
We report \emph{ab initio} fully relativistic calculations on the complex open-shell Sn$^{12+}$ highly charged ion. 
Previous calculations of the atomic structure of tin plasmas have consistently led to spectra that are significantly blue-shifted relative to experimental observations, well away from the 2\% reflectivity bandwidth that sets the minimum required accuracy for such calculations. We show that the complete CI saturation of the $n=4$ shell and the thorough inclusion of core-correlation, account for the crucial many-body correlations that explain these persistent discrepancies. A surprisingly large influence of highly-excited states on the energy levels contributing to the EUV spectrum is found. This originates from the collective effect of the configuration mixing of an extremely large number of those highly-excited states on the multiply-excited states that are involved in EUV transitions.

These insights directly translate to calculations of the other relevant in-band EUV emitting species, Sn$^{11+}$--Sn$^{14+}$. It is an open question whether transitions from multiply excited states are similarly dominant in heavier ionic systems such as, in particular, quasi-continuous emitters such as Gd, Tb ($\sim$6.7\,nm)~\cite{Kilbane2013}, and Bi (2.3--4.4\,nm)~\cite{Liu2017}. Again, their emission primarily involves open-shell $n=4-4$ transitions~\cite{Ohashi2014}, and the method described in this paper can unequivocally answer the question of the role of multiply excited states across a wide range of ionic species.

\textit{Acknowledgments}.---
This work was conducted at the Advanced Research Center for Nanolithography (ARCNL), a public-private partnership between the University of Amsterdam (UvA), Vrije Universiteit Amsterdam (VU), Rijksuniversiteit Groningen (UG), the Dutch Research Council (NWO), and the semiconductor equipment manufacturer ASML. 
This work used the Dutch national e-infrastructure with the support of the SURF Cooperative using grant no. EINF-13731. 
This publication is part of the project ARIES with file number 20152 of the research programme VENI which is financed by NWO. This work was funded by ERC CoG MOORELIGHT 101086839. 
EVK's work was supported by resources provided by the Pawsey Supercomputing Research Centre’s Setonix Supercomputer~\cite{Setonix2023} and PULSE research software engineering collaboration, with funding from the Australian Government and the Government of Western Australia. 
EVK thanks Sam Yates for useful discussions and advice on computational performance engineering. 
JS and OOV thank James Colgan and Amanda Neukirch for productive conversations.

\bibliography{references} 


\end{document}